# HYDRODYNAMICAL STUDIES OF WIND ACCRETION ONTO COMPACT OBJECTS: TWO-DIMENSIONAL CALCULATIONS

Benensohn, J. S.
*Department of Astronomy and Astrophysics*
*University of Chicago, Chicago, IL 60637*
*jeffb@oddjob.uchicago.edu*

Lamb, D. Q.
*Department of Astronomy and Astrophysics*
*University of Chicago, Chicago, IL 60637*
*lamb@oddjob.uchicago.edu*

Taam, R. E.
*Department of Physics and Astronomy*
*Northwestern University, Evanston, IL 60208*
*taam@ossenu.astro.nwu.edu*

## ABSTRACT

We present the results of hydrodynamical simulations of nonaxisymmetric gas flow past a gravitating compact object in two dimensions. Calculations were performed with uniform flow as well as with transverse velocity and density gradients. We find that the flow is highly nonsteady, exhibiting the "flip-flop" behavior seen in previous studies in which accretion disks form with alternating directions of rotation. We investigate the periodicity of the flip-flop behavior, and study the effects of spatial resolution on the results. We find that the flip-flop motion creates accretion torques which, in some cases, may be large enough to explain the erratic spin behavior observed in Be-type X-ray pulsars.

*Subject headings:* hydrodynamics – stars: accretion – stars: winds – X-rays: binaries

## 1. INTRODUCTION

Our understanding of accretion from a stellar wind onto a compact object has grown considerably in recent years through both observational and theoretical advances. Yet we still cannot explain much of the observations, especially in the case of magnetic neutron stars which are accreting from the stellar wind of a massive O/B-type companion star. These sources show intrinsically different spin behavior from magnetic neutron stars which accrete from a low-mass companion star that is overflowing its Roche lobe or from a Be-type star. In the case of Roche lobe overflow, an accretion disk forms due to the large angular momentum of the material passing through the inner Lagrangian point. The system is visible as a bright X-ray pulsar (see e.g. Lamb, Pethick, & Pines 1973), which is expected to spin up over long periods of time (Ghosh & Lamb 1979). In the case of a Be star, mass is periodically shed in the equatorial plane. The system is seen as a bright transient X-ray source (Stella, White, & Rosner 1986; van den Heuvel & Rappaport 1987). In this case the accreting material is expected to form a disk (Elsner, Ghosh, & Lamb 1980), so that the pulsar spins up during an outburst and spins down otherwise.

In contrast, it is not clear what to expect for the many X-ray pulsars in which a neutron star accretes from the wind of an O/B-type companion. Some of these sources show fluctuations in their spin period on timescales of hours to months (Boynton et al. 1984; Nagase et al. 1984; Nagase 1989; Chakrabarty et al. 1993). Many of them also show long-term trends of spin-up and spin-down of the neutron star (Nagase 1989; Chakrabarty et al. 1993, 1995; Chakrabarty 1996). However, there appears to be no characteristic timescale for the switch between spin-up and spin-down: observed timescales range from hours to years.

The first studies of the hydrodynamics of accretion from a moving ambient medium are the analytic treatments of Hoyle and Lyttleton (1939, 1940a, 1940b, 1940c). They neglect pressure effects, allowing the matter to focus gravitationally until it collides with itself behind the accreting object in an infinitely thin "accretion line." In the accretion line, the material is assumed to lose its transverse momentum. There is therefore a stagnation point at some distance downstream along the accretion line; material inside this point will accrete while material outside escapes downstream. With this simple picture, the mass accretion rate is

$$\dot{M}_{HL} = \pi R_a^2 \rho_\infty v_\infty , \qquad (1)$$

where the effective accretion radius is given by

$$R_a = 2GM/v_\infty^2 . \qquad (2)$$

Here M is the mass of the accreting object, and $\rho_\infty$ and $v_\infty$ are the density and velocity of the medium at infinity upstream from the accreting object. Bondi (1952) considered the case of accretion from a static ambient medium of finite temperature, and suggested the following interpolation formula:

$$\dot{M}_{BH} = 0.5 \left( \frac{\mathrm{M}_\infty}{\mathrm{M}_\infty + 1} \right)^{3/2} \dot{M}_{HL} , \qquad (3)$$

where $M_\infty$ is the Mach number at infinity (although most authors now omit the factor of 0.5 in eq. [3]).

Application of the Bondi-Hoyle-Lyttleton model to accretion by a compact object from the stellar wind of a binary companion star has been somewhat controversial. In a binary system, the orbital motion introduces transverse gradients in density and velocity as seen by the compact object. This results in a net amount of angular momentum deposition in the accretion cylinder (Illarionov & Sunyaev 1975; Shapiro & Lightman 1976; Anzer, Borner, & Monaghan 1987). However, the analytic Bondi-Hoyle-Lyttleton approach treats the accretor as a point mass and the accretion wake as an infinitely thin line; consequently, the angular momentum accreted is always exactly zero, even if a transverse gradient exists (Davies & Pringle 1980). Hence it is unclear what fraction of the angular momentum deposited in the accretion cylinder will be accreted by the central object, and how this might occur. Pressure effects make the accretion process much more complex than this simplest picture, particularly in high Mach number flows wherein shocks and turbulent motions are important. No analytic model can incorporate these effects properly, so numerical calculations are our only means of modelling the process of accretion onto a compact object from a stellar wind.

Beginning over forty years ago, many authors have examined the validity of the Bondi-Hoyle-Lyttleton analytic estimates of the mass accretion rate (eqs. [I.1] and [I.3]) using hydrodynamical calculations (Dodd 1952, 1953; Dodd & McCrea 1952; Hunt 1971, 1979; Shima et al. 1985; Anzer, Borner, & Monaghan 1987). These calculations showed that the estimate of the mass accretion rate in equation (I.1) is quite good, despite the simplicity of the Bondi-Hoyle-Lyttleton picture. However, in these early calculations either the flow was assumed to be axisymmetric or the numerical resolution was low, and only steady-state solutions were found.

As advancing computer power permitted, higher-resolution non-axisymmetric hydrodynamical calculations were performed (Matsuda, Inoue, & Sawada 1987, Fryxell & Taam 1988, Taam & Fryxell 1988,

Taam & Fryxell 1989, Matsuda et al. 1991). These studies revealed that the accretion process is highly nonsteady: at moderate Mach numbers, the tail shock behind the compact accreting object is unstable to tangential oscillations. This so-called "flip-flop" instability has an interesting consequence in accretion flows: it leads to the formation of disk-like rotational inflows which alternate in direction. The compact object accretes substantial amounts of angular momentum while the disk-like inflows are present, and the mass accretion rate shows variations of up to a factor of ten. The instability may be similar in origin to the shedding of von Karman vortex "streets" observed in high Reynolds number flow past a hard cylinder. For a review of the calculations to that time see Livio (1992); for a review of the von Karmann vortex street phenomenon see Williamson (1996).

The flip-flop instability may account for the periodic flaring behavior seen in some transient Be-type X-ray binaries: the calculated mass accretion rates bear a strong resemblance to the observed X-ray flux (Taam, Fryxell, & Brown 1988). In such cases, since the mass outflow from the Be-type companion star is thought to be confined to the equatorial plane, two-dimensional calculations may represent a fair approximation of the system. However, two-dimensional calculations are inadequate to model O/B-type high-mass X-ray binaries, wherein the stellar wind outflow is spherical; for such situations three-dimensional calculations are required.

Early three-dimensional calculations were performed by Livio et al. (1986), Matsuda et al. (1992), and Ishii et al. (1993), finding generally stable Mach cones and no flip-flop instability, but the spatial resolution attained in these calculations was too low to draw any firm conclusions. The first high-resolution three-dimensional calculations were the pioneering studies of Ruffert (1994a, 1994b, 1995, 1996). Ruffert's calculations of uniform flow past a gravitating sphere show a roughly steady bow shock with a turbulent wake, but no "flip-flop" instability. However, his most recent calculations (Ruffert & Anzer 1995, Ruffert 1996b) which include transverse velocity gradients show strong nonsteady behavior, including the occasional formation of small rotational inflows. These disk-like inflows are smaller and less ubiquitous than in the two-dimensional calculations; they do not appear at regular intervals or alternate in direction the way they do in two dimensional calculations. Fluctuations are seen in the mass accretion rate, but their amplitude is only tens of percent, as opposed to the variations of a factor of ten seen in two-dimensional calculations. More high-resolution three-dimensional calculations which include transverse density gradients would be useful to help resolve this issue. Results of such calculations will be presented in a future paper (Benensohn 1996).

In this paper we present results from extensive two-dimensional hydrodynamical calculations, including a study of the effects of resolution on the results. In section 2 we describe our computational method; in section 3 we present our results; in section 4 we discuss our results and describe our resolution study; in section 5 we summarize our conclusions.

## 2. COMPUTATIONAL METHOD

We formulate the problem as follows: an ambient medium flows supersonically past a gravitating object. We use an Eulerian formulation with a fixed grid of cells in cylindrical coordinates $(r, \phi)$. The gravitating object is placed at the center of the coordinate system ($r = 0$), so that the surface of the object (corresponding to the magnetosphere of the neutron star) is the inner boundary of the grid. The grid thus extends from $r_{\min}$ to $r_{\max}$, and $r_{\min}$ is the radius of the central object. The computational cells are divided uniformly in azimuthal angle, so that the angular extent of each cell is $2\pi/N_\phi$ where $N_\phi$ is the number of azimuthal zones. The sizes of radial zones are assigned according to

$$\Delta r_{i+1} = \beta \Delta r_i , \qquad (4)$$

with $\beta > 1$ so that the radial size of a cell increases with distance from the central object. The values of $\beta$ and $\Delta r_1$ determine both the fineness of the grid's radial resolution and the maximum extent of the grid $r_{max}$, and can be different in different calculations. Usually $\beta$ and $\Delta r_1$ are chosen so that cells everywhere in the grid remain roughly square, i.e. $\Delta r_i \approx r(i)\Delta\phi$. We additionally require $\beta \leq 1.05$ to avoid errors in the numerical integrations.

The grid is two-dimensional, yet we use a three-dimensional gravitational potential; this corresponds to the assumption that the central object moves through a thin sheet of ambient material. The relative velocity $v_\infty$ is fixed at infinity along $\phi = 0$. The fundamental units in the problem are the sound speed at infinity $c_\infty$ and the accretion radius $R_a$. For all cases considered in this paper, we take a $1.4 M_\odot$ central object, and at infinity along $\phi = 0$ we set the velocity to 1000 km s$^{-1}$ and the Mach number to 4, giving an accretion radius of size $3.7 \times 10^{10}$cm.

Calculations are performed using the Piecewise Parabolic Method (Colella & Woodward 1984), neglecting self-gravity of the gas, radiative heating and cooling, and magnetic effects. The gas follows an adiabatic equation of state with $\gamma = 4/3$.

The inner boundary of the grid, $r = r_{min}$, corresponds to the surface of the magnetosphere of a neutron star. We therefore use a totally absorbing inner boundary, accomplished by maintaining a very small density and pressure immediately inside the boundary. Consequently all mass and angular momentum reaching this magnetospheric radius is accreted, yielding the maximum possible accretion rates. All calculations in this paper have $r_{min} = 0.0375 R_a = 1.4 \times 10^9$cm. This size is somewhat larger than expected for the magnetospheric radius of a strongly magnetic neutron star in disk accretion, which is given by

$$r_0 \approx 0.5 r_A = 1.8 \times 10^8 L_{37} B_{12}^{4/7} \text{ cm} \tag{5}$$

for a neutron star with mass $1.4 M_\odot$ and radius 10 km, where $L_{37}$ is the X-ray luminosity in units $10^{37}$ erg s$^{-1}$, $B_{12}$ is the magnetic field in units $10^{12}$ Gauss, and $r_A$ is the characteristic Alfvén radius for spherical accretion (see Ghosh & Lamb 1979 and *e.g.* eq. [15.1.6] in Shapiro & Teukolsky 1983). However, we have chosen this radius for the central object for computational considerations and in order to allow more direct comparison with previous calculations that used central objects of this size.

The outer boundary of the grid, $r = r_{max}$, is divided into two parts: upstream and downstream. Downstream from the central object ($\pi/2 < \phi < 3\pi/2$), we allow material to flow freely off of the computational grid. Upstream from the central object ($-\pi/2 < \phi < \pi/2$), the parameters of the material (density, velocity, temperature) are fixed at a distance of $r = \infty$. In some cases, a transverse gradient in density or velocity is imposed at $r = \infty$ upstream from the accreting object. Such gradients are the result of the superposition of the accretor's orbital velocity and its companion's wind velocity (Shapiro & Lightman 1976; Anzer, Borner, & Monaghan 1987; Ho 1988). A two-dimensional analysis similar to that done by Shapiro & Lightman (1976), which assumes a constant wind velocity with radius, shows that the expected size of the difference in density across one accretion radius $R_a$ is, to leading order in $v_{orb}/v_w$,

$$\frac{\Delta\rho}{\rho} \approx \frac{2 R_a v_{orb}}{D\sqrt{v_{orb}^2 + v_w^2}}, \tag{6}$$

where $v_{orb}$ is the orbital velocity, $v_w$ is the wind velocity, and $D$ is the binary separation. Similarly, the expected velocity difference across one accretion radius is, to leading order in $v_{orb}/v_w$,

$$\frac{\Delta v}{v} \approx -\left(\frac{v_{orb}}{v_w}\right)^2 \times \left(\frac{R_a v_{orb}}{D\sqrt{v_{orb}^2 + v_w^2}}\right). \tag{7}$$

Hence the velocity difference is smaller by a factor of $(v_{orb}/v_w)^2$ than the density difference. Typical values of this factor in high-mass OB supergiant systems range from 0.1 to 0.2; in Be-binary systems typical values are less than 0.1. Hence in such systems the density gradient is the dominant effect of the orbital motion. Table 1 shows the expected values of $\epsilon_\rho$ and $\epsilon_v$ in some well-known high-mass X-ray binary systems. The estimates have been made using equations (6) and (7), assuming a circular orbit, and taking the stellar wind speed $v_\infty$ to be the escape speed at the star's surface given its spectral type. Values for $\epsilon_\rho$ range from $6.1 \times 10^{-4}$ to $2.3 \times 10^{-2}$, with Be-type systems having values below $1.6 \times 10^{-3}$. However, many Be-type systems are known to have highly noncircular orbits with outbursts occuring at periastron; hence during outburst the higher orbital velocity and smaller orbital distance will produce correspondingly larger transverse gradients. Note that this analysis neglects rotation of the giant companion star; if the companion star is rotating rapidly and the binary separation is very small, the transverse gradients can be somewhat smaller, although in most cases this contribution is negligible (Illarionov & Sunyaev 1975).

We impose these gradients by assuming a form for the density and velocity at $r = \infty$ upstream from the accreting object of

$$\rho = \rho_\infty (1 + \epsilon_\rho r \sin\phi / R_a) \qquad (8)$$

and

$$v = v_\infty (1 + \epsilon_v r \sin\phi / R_a) \ . \qquad (9)$$

When a density gradient is imposed, a gradient in pressure is also imposed according to $P \propto \rho^\gamma$ in order to remain consistent with our assumption that the fluid has a constant ratio of specific heats $\gamma$. A density gradient therefore also implies a gradient in the temperature (as $T \propto \rho^{\gamma-1}$) and in the sound speed (as $c \propto \rho^{1/2(\gamma-1)}$), but both of these gradients are very small as we use $\gamma = 4/3$.

After fixing the wind parameters at infinity, we use a ballistic approximation developed by Bisnovatyi-Kogan et al. (1979) to project the flow from infinity onto the edges of the computational grid (and onto the entire upstream grid for initial conditions). This approximation gives an extremely accurate representation of the upstream flow for $r > 2R_a$ and Mach numbers greater than 1.4 (Koide, Matsuda, & Shima 1991). We use a Mach number of 4 and never impose an outer boundary closer than $10R_a$.

Most previous studies simply impose the infinite-distance values on the upstream boundary. We find that imposing the infinite-distance values of the flow at an outer radius of $5R_a$ will underestimate the velocity of the flow at this outer radius by 10%, while imposing the uniform flow at $10R_a$ underestimates the velocity by 5%; these results are in agreement with Koide, Matsuda, & Shima (1991). Additionally, we find that a transverse gradient imposed on the flow at $5R_a$ underestimates the value of $\epsilon$, the size of the gradient, by 10% due to the gravitational focussing of the material from infinity. Hence a calculation which imposes infinite-distance boundary conditions at a finite radius is actually using an effectively lower Mach number and effectively smaller transverse gradients. Since all studies to date show the flow to depend strongly on the values of the Mach number and the transverse gradient, these effects can become important for evaluating differences between calculations and for drawing conclusions about real X-ray binary systems.

Given the transverse gradients at infinity, one can calculate the amount of specific angular momentum deposited within an accretion radius (the total angular momentum deposition rate divided by the mass deposition rate within an impact parameter of one accretion radius):

$$j = \frac{\dot{J}}{\dot{M}} = \frac{1}{3}(\epsilon_\rho - 6\epsilon_v)v_\infty R_a \ . \qquad (10)$$

In three dimensions the 1/3 factor becomes 1/4. This equation is exact for $\epsilon_\rho$ but is only a first-order approximation for $\epsilon_v$, requiring $\epsilon_v \ll 1$. For the velocity gradient a simple analytic formula is not possible because the accretion radius also varies as the velocity varies, extending the effective accretion radius on the low-velocity side and contracting it on the high-velocity side (making the face of the accretion cylinder noncircular). This effect is the cause of the two gradients having opposite sign: a velocity gradient deposits angular momentum in the opposite sense of the density gradient because the extension of the accretion radius on the low-velocity side is more important than the increase in velocity on the high-velocity side. In fact if $\epsilon_v > 0.15$, the effective accretion radius on the low-velocity side is infinite. In practice, values of $\epsilon_v$ which are greater than $\approx 0.1$ require a cutoff in the extent of the gradient both to prevent an infinite accretion radius and to prevent negative flow velocities in the stellar wind. Since the density gradient is expected to be the dominant effect in typical high-mass X-ray binaries, only one calculation in this paper uses a velocity gradient (of size $\epsilon_v = 0.005$).

## 3. RESULTS

In this section we present results from six different calculations. Table 2 summarizes the calculations and parameters used. In subsection 1, we investigate uniform flow using three different spatial resolutions; in subsection 2, we investigate a transverse velocity gradient; in subsection 3, we investigate two transverse density gradients.

### 3.1. Uniform Flow

In this section we present results from three different calculations (using low, medium, and high spatial resolution) in which the stellar wind material is assumed to be uniform in density and velocity. In all three cases we apply a uniform medium moving with speed $1000$ km s$^{-1}$ = Mach 4 at $r = \infty$ upstream in the $\phi = 0$ direction.

### 3.1.1. Low Resolution

For this calculation we use a low-resolution grid, with 100 uniformly spaced azimuthal zones and 200 nonuniformly spaced radial zones. The radial zones are gridded according to equation (4), with $\Delta r_1 = 0.14 r_{min} = 0.00525 R_a$ and $\beta = 1.02$. The outer boundary therefore falls at $373 r_{min} = 14.0 R_a$.

In this calculation the low resolution suppresses the growth of the "flip-flop" instability; it does develop, but not until $t \approx 20$. Once it finally reaches its maximum amplitude, its behavior is similar to that seen in previous studies: an accretion tailshock develops and becomes unstable, waving from side to side downstream of the accreting object. As the wake swings to one side, an accretion disk forms around the object; when the wake swings back towards the other side, the disk dumps its matter rapidly onto the object and a new disk forms rotating in the opposite direction.

The accretion rates are presented in Figure 1. The mass accretion rate is normalized to the Hoyle-Lyttleton analytic rate (eq. [1]). The specific angular momentum of the accreted material is given in units of $R_a c_\infty$; in these units, Keplerian rotation at the surface of the accretor corresponds to a value of 0.56. The time units are $R_a/c_\infty = 25$ minutes. The effects of the "flip-flop" instability are apparent as the specific angular momentum accretion rate oscillates between its positive and negative extrema, the values of which indicate slightly larger than Keplerian rotation at the inner boundary. The mass accretion rate is small when a disk is present and shows large spikes as each accretion disk dumps onto the object. The average mass accretion rate is $0.937 \dot{M}_{HL}$, and the mean period of the oscillation is 1.89 time units. The total angular momentum captured oscillates in a manner consistent with a random walk. There is no evidence for a long-term net trend in the accretion of angular momentum despite the large short-term variations.

### 3.1.2. Moderate Resolution

For this calculation we use a moderate-resolution grid, with 224 uniformly spaced azimuthal zones and 224 nonuniformly spaced radial zones. The radial zones are gridded according to equation (4), with $\Delta r_1 = 0.027 r_{min} = 0.001 R_a = 3.7 \times 10^7$ cm and $\beta = 1.025$. The outer boundary therefore falls at $296 r_{min} = 11.1 R_a = 4.1 \times 10^{11}$ cm.

In this calculation, the "flip-flop" instability develops immediately and reaches maximum amplitude by $t \approx 3$. Figure 2 shows a grayscale representation of the density in the flow: the compact object lies at (0,0), and the stellar wind is incident from the right. The unstable side-to-side motion of the wake creates a snaking back-and-forth pattern downstream of the accreting object. The pattern is strongly reminiscent of von Karmann vortex street shedding (see *e.g.* Williamson 1996) and is likely similar in origin, as both are created by turbulent shear layers in flow past a cylinder. However, a detailed comparison of the two behaviors is not applicable, as the presence of the strong gravitational field and an accreting inner boundary are unique to the flip-flop instability.

The accretion rates are presented in Figure 3. The average mass accretion rate is $0.866 \dot{M}_{HL}$, and the mean period of the oscillation is 1.98 time units. The total angular momentum captured oscillates in a manner similar to a random walk. There is no evidence for a net trend in the accretion of angular momentum.

Figure 4 shows close-up views of an accretion disk in this calculation. (a) shows a grayscale representation of the density at time t=8.09, and part (b) shows the velocity field at this time. The radius of this disk is at least $0.27 R_a$, within which radius the grid has 75 radial zones, so this accretion disk is well resolved. This disk is a fairly representative picture of the maximum size reached by each accretion disk which forms. However, it is not representative of the way the disk looks for a long period of time. The accretion disks are highly nonsteady, fluctuating significantly in size and shape during each disk phase. This variability is evident to some extent in the accretion rates (Figure 3.2), but is more obvious in a time sequence of the density in the disk. Consider Figure 4(c), which shows the density in the accretion disk at t=8.07; the disk looks similar but the detailed structure of the spiral shocks is different. Figure 4(d) shows the same accretion disk at t=8.02, and the disk is seen to look very different; the disk is smaller and is noncircular. As these images indicate, the shape of the disk is constantly changing, but in general the disk has radius $0.1 R_a - 0.3 R_a$ and two main spiral shocks with complicated substructure.

Hydrodynamic shocks are the primary means of outward angular momentum transfer in these disks, and they are relatively inefficient. Comparing the disk in Figure 6 with an alpha-disk model (Shakura & Sunyaev 1973), we find effective values of $\alpha \approx 10^{-4}$ for annulus-averaged quantities, although individual computational cells show maximum values of $\alpha \approx 0.09$. Note however that the alpha-disk model assumes smoothly varying quantities with radius and a fixed viscous drag between each thin annulus; hence it does not directly apply to disks with strong spiral shocks like those seen in these calculations.

### 3.1.3. High Resolution

For this calculation we use a high-resolution grid, with 320 uniformly spaced azimuthal zones and 272 nonuniformly spaced radial zones. The radial zones are gridded according to equation (4), with $\Delta r_1 = 0.027 r_{min} = 0.001 R_a = 3.7 \times 10^7$ cm and $\beta = 1.02$. The outer boundary falls at $315 r_{min} = 11.8 R_a$.

The accretion rates are presented in Figure 5. We find that the behavior at this resolution is qualitatively and quantitatively the same as that seen in lower resolution calculations. The average mass accretion rate is $0.845 \dot{M}_{HL}$. The average period of oscillations is 2.28 time units, somewhat longer than that at lower resolution, but this is not significant as the calculation has only been integrated for four flip-flop periods, and the last period seen is only 1.7 time units. The total integrated angular momentum captured oscillates in a manner similar to a random walk; the final mean value happens to be negative only because the first disk formed rotating in a clockwise direction. There is no evidence for a net trend in the accretion of angular momentum.

Accretion disks in this calculation show typical maximum sizes of $0.25 - 0.30 R_a$, consistent with those seen in §3.1.2 above. Comparison with an alpha-disk model gives effective values of $\alpha \approx 10^{-4}$ for annulus-averaged quantities, although individual computational cells show maximum values of $\alpha \approx 0.06$. These values are consistent with those seen in §3.1.2. However, as mentioned in §3.1.2 above, alpha-disk models are based on assumptions which do not fully apply to the disks seen in these calculations.

### 3.2. Transverse Velocity Gradient

This calculation is identical to that in §3.1.2, with the same moderate-resolution grid, except that we now impose a transverse velocity gradient on the flow at infinity. The size of the gradient is $\epsilon_v = 0.005$ (see eq. [9]). While this is larger than the value expected in typical Be-type X-ray binaries (see Table 1), we wish to compare the effects of velocity gradients to those of density gradients, which are presented later in this paper.

The accretion rates are presented in Figure 6. They display the same qualitative behavior as in the uniform flow: an unstable accretion wake and disks forming with alternating directions. The average mass accretion rate for this case is $0.834 \dot{M}_{HL}$, and the mean period of the oscillations is 1.96 time units. The total angular momentum captured oscillates in a manner consistent with a random walk; the mean value happens to be negative only because the first disk formed rotating in a clockwise direction. There is no evidence for a net trend in the accretion of angular momentum over time.

### 3.3. Transverse Density Gradients

### 3.3.1. Small Density Gradient

This calculation is identical to that in §3.1.2, with the same moderate-resolution grid, except that we now impose a transverse density gradient on the flow at infinity. The size of the gradient is $\epsilon_\rho = 0.005$ (see eq. [8]), a value expected in typical high-mass X-ray binary systems (see Table 1).

The accretion rates are presented in Figure 7. They display the same qualitative behavior as in the uniform flow: an unstable accretion wake and disks forming with alternating directions. The average mass accretion rate is $0.875 \dot{M}_{HL}$, and the average period of the oscillations is 1.96 time units. The extreme length of this calculation reveals a new and interesting behavior: after an initial phase during which the total angular momentum accreted ($J_{tot}$) increases with time, the trend reverses, temporarily producing a small negative slope, then seems to reverse yet again. We therefore see the possibility of three timescales in $J_{tot}$: short-term reversals caused by the flip-flop instability, medium-term trends wherein $J_{tot}$ shows net increases or decreases over 5-10 flip-flop periods, and perhaps a long-term trend wherein the mean value of $J_{tot}$ slowly increases due to the small density asymmetry in the incident medium. Accurate statistical analysis of the medium-term and long-term trends would require the calculation to be integrated for much longer, but after 31 flow timescales we now have enough evidence to suggest the longer-term behavior. We conjecture that the long-term behavior of $J_{tot}$ will be characterized by a general long-term positive trend with some periods showing medium-term downward trends.

### 3.3.2. Moderate Density Gradient

This calculation is identical to that in §3.3.1 above, with the same moderate-resolution grid, except that we increase the size of the transverse density gradient by a factor of 5 to $\epsilon_\rho = 0.025$, both to investigate the dependence of our results on the size of the gradient and to model high-mass X-ray binary systems with shorter orbital periods (see Table 1).

The accretion rates are presented in Figure 8. They display the same qualitative behavior as in the uniform flow: an unstable accretion wake forms, leading to a succession of disks rotating in alternate directions. The average mass accretion rate is $0.864\dot{M}_{HL}$, and the average period of the oscillations is 1.96 time units. The time-integrated angular momentum has a small positive slope of $\approx 2.5$, indicating a small positive long-term trend. The calculation does not extend long enough to show the medium-term trend reversals seen in §3.3.1 above. It is clear that larger gradients than $\epsilon_\rho = 0.025$ are required to show the strong continuous trends in total angular momentum accreted like those seen with $\epsilon_\rho = 1.0$ in Fryxell & Taam (1988).

## 4. DISCUSSION

### 4.1. Comparison with Previous Work

The calculations presented in this paper were designed to be similar to those of Taam & Fryxell (1988, 1989) and Fryxell & Taam (1988). Those calculations have the same Mach number (4), a similar wind speed (700 km s$^{-1}$), a similar central object (mass $1 M_\odot$ and radius $0.037 R_a$). Their transverse gradients are implemented in the same way; they use $\epsilon_\rho$ values of 0.005, 0.0625, 0.25, and 1.0, and $\epsilon_v$ values of 0.005 and 0.0625. Their numerical method is nearly identical: they use the Piecewise Parabolic Method with a constant ratio of specific heats $\gamma = 4/3$ on a fixed cylindrical grid. The spatial resolution of their grid is not as fine, but it is adequate for the calculations (a detailed discussion of numerical resolution is given in §4.4).

As might be expected, our results are very similar to the results of Taam and Fryxell. For example, compare our Figure 7 to Figures 1, 7, and 14 in Fryxell & Taam (1988); the mass accretion rate, specific angular momentum, and total time-integrated angular momentum are strikingly similar. One noticeable difference is that in our calculations there are more small fluctuations in the mass accretion rate; we can resolve these fluctuations due to our finer spatial resolution. Another important difference is that we see a small increasing trend in angular momentum accretion with $\epsilon_\rho = 0.005$, whereas they do not. The only other noticeable difference is that we integrate the flow for 15 flip-flop periods, twice as long, allowing a more detailed investigation of long-term trends in angular momentum accretion.

Most other previous two-dimensional calculations (*e.g.*, Ishii et al. 1993) suffer from a lack of spatial resolution and are badly affected by numerical effects; we give a detailed discussion of those calculations in §4.4.

### 4.2. Periodicity of the Flip-Flop Instability

The nature of the flip-flop instability is more fully understood when one investigates the periodicity of the back-and-forth motions which it causes. Figure 9 shows the power spectral density (PSD) of the specific angular momentum accretion rate for two different calculations. The upper panel shows the PSD from the calculation presented in §3.3.1 (small density gradient, moderate resolution grid). The lower panel shows the PSD from the calculation presented in §3.1.1 (uniform flow, low resolution grid). Each plot displays the absolute magnitude of the Fourier transform of the specific angular momentum accreted (in arbitrary units) versus frequency (in units $c_\infty/R_a \approx 0.67$ mHz).

The periodic behavior of the flow causes the peak at low frequency, and the power at high frequencies is characterized by a power law with slope $\approx -0.5$. The high frequency power is not noise; it is caused by real fluctuations in the accretion rate due to physical fluid motions. For the power spectral density plot in the upper panel of Figure 9, the peak in the power spectrum has $f_0/\delta f \approx 2.5$. Among all calculations, we find a range of $f_0/\delta f \approx 1 - 5$. Many people require $f_0/\delta f > 5$ in order for a phenomenon to be quasiperiodic (see, *e.g.*, van der Klis 1995). According to this criterion, the flip-flop motions are not quasiperiodic.

### 4.3. Implications for X-Ray Pulsars

Long-term trends in the time-integrated angular momentum accretion rate like those seen in §3.3.1 can cause secular increases or decreases in the spin period of the accreting object. The accretion torque creates spin changes via $I\dot\Omega = \dot J$, where $I$ is the object's moment of inertia and $\Omega$ is the object's spin frequency; hence

$$\frac{\dot P}{P} = -\frac{P\dot J}{2\pi I} \,. \tag{11}$$

Consider a Be-type X-ray pulsar in outburst with pulse period 100 sec and X-ray accretion luminosity $10^{37}$ erg s$^{-1}$ (corresponding to a mass accretion rate of $10^{17}$ $rmg$ s$^{-1}$), and assume a neutron star mass of $1.4 M_\odot$ and radius of 10 km. Taking an average slope from Figure III.7 of $\dot J_{tot} \approx 0.27$, we find $\dot P/P \approx (2 \times 10^{-4}) L_{37}$ yr$^{-1}$, where $L_{37}$ is the accretion luminosity in units $10^{37}$ erg s$^{-1}$. During disk phases, slopes become as large $\dot J_{tot} \approx 4$, leading to $\dot P/P \approx (3 \times 10^{-3}) L_{37}$ yr$^{-1}$. Spin-period changes in Be-type systems during outburst have been observed as large as $\dot P/P \approx (10^{-2})$ yr$^{-1}$ with long-term average rates of $\dot P/P \approx (10^{-3})$ yr$^{-1}$ (see *e.g.* Nagase et al. 1982, Nagase et al. 1989, and Parmar et al. 1989). In order to produce spinup/spindown rates of this size in the calculations, X-ray accretion luminosities of $L \approx 4 \times 10^{37}$ erg s$^{-1}$ are required. The calculations also predict a spinup/spindown rate which is proportional to the X-ray luminosity; this effect is also observed (Nagase et al. 1982). It therefore appears that the erratic spin behavior of some Be-type X-ray pulsars can be explained by the flip-flop instability if the accretion luminosity reaches $L \approx 4 \times 10^{37}$ erg s$^{-1}$.

The magnitude of $\dot P/P$ obtained in these calculations is due in large part to the lever arm obtained by the accreting material: the angular momentum is accreted at the surface of the magnetosphere, yet it goes towards the spinup of a very compact star. In all of the Mach 4 calculations in this chapter, the magnetospheric surface is at $1.4 \times 10^9$ cm, yet the neutron star's radius is a mere 10 km. This lever arm, which provides a factor of 1000 in $\dot P/P$, requires a strong magnetic field of size $B \geq 3 \times 10^{12}$ Gauss for an accretion luminosity of $4 \times 10^{37}$ erg s$^{-1}$ (see eq. [III.7]). Pulsars with weaker magnetic fields will have a smaller lever arm for accretion of angular momentum and will produce correspondingly smaller values of $\dot P/P$. This consideration is an important caveat to the above discussion, making it somewhat more difficult to account for all spin behavior observed in high-mass X-ray binary systems via the flip-flop instability.

The comparison of our results to observations are hampered somewhat by a difference in timescale: the calculations cover shorter time periods (several hours) than have been observationally probed. Perhaps with the recent launch of the Rossi X-ray Timing Explorer, new observations on short timescales will reveal fluctuations in the spin period of some X-ray pulsars which could be explained by these calculations.

### 4.4. Numerical Resolution

Most numerical studies of wind accretion done to date suffer from the same problem: they are limited by the availability of computer resources. Most hydrodynamical calculations simply use the maximum spatial resolution possible given the available computer power. Unfortunately, this precludes determining whether or not the spatial resolution is adequate. As Fryxell (1994) has demonstrated, thorough resolution studies sometimes reveal that previous results were affected by inadequate spatial resolution.

We have performed many calculations with different grid sizes in order to investigate the effects of numerical resolution on the results. Tests were performed using different numbers of uniformly spaced azimuthal zones (100, 128, 160, 200, 224, and 320). The radial zones were arranged with $\Delta r_{i+1} = \beta \Delta r_i$, with $0 < \beta < 1.05$; the specific value of $\beta$ is usually chosen to match the angular resolution (so that each zone remains roughly square). Tests were also performed with 224 azimuthal zones using different radial resolutions. Results from these test calculations were compared to look for both qualitative and quantitative differences in the results. One useful comparison tool we employ is the power spectral density of the mass and angular momentum accretion rates (e.g., Figure 9). We find that inadequate resolution in either coordinate direction has noticeable effects on the flow.

First, inadequate resolution suppresses the onset of the instability. With 100 azimuthal zones, the instability develops very slowly in uniform flow, not reaching its maximum amplitude until $t \approx 20$. With 200 azimuthal zones or more, the instability develops immediately, and the maximum amplitude is reached before $t = 3$. Large transverse gradients can start the instability more quickly by breaking the initial axisymmetry.

Second, inadequate resolution causes errors in the average mass accretion rate. With 100 azimuthal zones, the average mass accretion rate is $0.937 \dot{M}_{HL}$. By contrast, every calculation we have performed with 200 azimuthal zones or more gives a value of $<\dot{M}>$ which is within 2.7% of the mean value $0.8568 \dot{M}_{HL}$.

Third, inadequate resolution suppresses high-frequency fluctuations in the flow. Considering the power spectral density plots in Figure 9, the PSD of the low-resolution calculation show a factor of two to three reduction of power at high frequencies as compared to the higher-resolution calculation. Additionally, in calculations done with 224 azimuthal zones, I find that poor radial resolution (with $\Delta r_1 = 0.07$ times the radius of the inner boundary) causes a suppression of power by a factor of $\approx 1.5$ at high frequencies. I find that with 224 azimuthal zones, using a finer radial gridding (with $\Delta r_1 = 0.027$ times the radius of the inner boundary) produced power spectra that are indistinguishable from those results using 320 azimuthal zones. This high-frequency power is not noise, but rather real fluctuations due to the motion of small structures in the flow. The results of this resolution study indicate that the minimum adequate resolution is $\approx 200$ azimuthal zones and a radial resolution which makes each cell roughly square; that is, $\Delta r_i \approx r(i) \Delta \phi$. Hence most of the calculations in this paper have 224 angular zones with extremely fine radial gridding.

Taam & Fryxell (1988, 1989) and Fryxell & Taam (1988) use 200 azimuthal zones and 100 radial zones (gridded with $\Delta r_1 = 0.25$ times the radius of the inner boundary and $\beta = 1.03$ or $1.05$). Our resolution study indicates that the spatial resolution they use is sufficient for this problem.

In contrast, Ishii et al. (1993) use a Cartesian grid with $64 \times 64$ cells, with a central accretor that is a $2 \times 2$ square, and Matsuda et al. (1991) use (at best) a Cartesian grid with $128 \times 128$ cells and a 1-cell square central object. Calculations we have done with similar low-resolution Cartesian grids indicate that they are inadequate to describe the flip-flop behavior either qualitatively or quantitatively. In particular, the use of a square central object that consists of only one or four cells introduces large errors in the angular momentum accretion rate and can lead to nonphysical effects. No accretion disk whatsoever is seen in the results of Ishii et al. (1993) or Matsuda et al. (1991), and the mass accretion rate does not show any fluctuation greater than 20%. High radial resolution, such as that used in the calculations presented in this paper, are required to better resolve the gravitational field (which changes rapidly near the central object) and reduce the numerical viscosity so that accretion disks may form more easily and persist longer.

Finally, we note that this resolution study was performed using specific values of the Mach number and inner boundary radius, and may not apply for all parameter space. For example, large central objects and/or low Mach numbers may lead to stable time-independent solutions, for which lower resolution would likely be adequate. Conversely, smaller central objects and/or higher Mach numbers may lead to a higher degree of turbulence, for which higher resoultion could be advantageous.

### 4.5. Correlation of Accretion Rates

A further insight into the nature of the flip-flop instability is provided by observing the correlation of the specific angular momentum of the accreted material (j) with the mass accretion rate ($\dot{M}$). Figure 10 shows scatter plots of j versus $\dot{M}$ for four of the six calculations presented in this paper; each dot represents a value averaged for 20 timesteps. All four plots reveal the same strong correlation: the mass accretion rate can only be significantly lower than the Hoyle-Lyttleton value when the specific angular momentum is near an extremum. This produces a characteristic rectangular shape in the plot. (The small feature near j=0 with small mass accretion rate is produced at the very beginning of the calculation.) This correlation provides another diagnostic for the flip-flop behavior which can be used for comparison purposes with future calculations.

## 5. CONCLUSIONS

We present the results of high-resolution two-dimensional hydrodynamical simulations of accretion by a compact object from a high-velocity ambient medium. We confirm that the "flip-flop" instability is a robust phenomenon in two dimensions that appears when material flows supersonically past a small accreting body with moderate Mach number. The instability produces recurring accretion disks that alternate in direction, although the recurrence time is not quite regular enough to be classified as quasiperiodic. While these disks are present the compact object accretes substantial amounts of angular momentum, roughly equivalent to Keplerian rotation at its surface and far greater than the amount deposited within an accretion radius by the flowing material. The instability occurs in uniform flow, wherein zero angular momentum is deposited by the flowing material.

The angular momentum accreted by the compact object as a result of the flip-flop instability creates accretion torques which can explain the erratic spin behavior observed in some Be-type X-ray pulsars. The instability causes both short-term and long-term trends and trend reversals in the spin period and spin period derivative of the compact object. To explain the magnitudes of $\dot{P}/P$ which are observed, high accretion luminosities ($L \approx 4 \times 10^{37}$ erg s$^{-1}$) and strong pulsar magnetic fields ($B \geq 3 \times 10^{12}$ Gauss) are required. Yet the flip-flop instability remains a viable model for the otherwise unexplainable erratic spin behavior observed in many Be-type X-ray pulsars and for the transient flaring activity observed in some Be-type X-ray pulsars.

Our resolution study indicates that this problem is not fully resolved with less than $\approx 200$ angular zones or with radial resolution violating $\Delta r_1 \leq 0.03$ times the radius of the inner boundary. Additionally, imposing boundary conditions at a finite radius without using a ballistic approximation leads to underestimates of both the Mach number of the flow and the size of the transverse gradients.

### Acknowledgements

This project would not have been possible without the support of Prof. Tom Prince of the California Institute of Technology, who generously provided computer time on the Intel Touchstone Delta supercomputer. JSB would like to thank Bruce Fryxell, Andrea Malagoli, and Paul Ricker for helpful conversations during this project. Research at the University of Chicago is supported in part by NASA grants NGT-51059, NAGW-830, NAGW-1284, and NAG 5-2868; research at Northwestern University is supported in part by NASA grant NAGW-2526.

# FIGURE CAPTIONS

Figure 1. Accretion rates in the calculation with uniform flow and low resolution. Time units are $R_a/c_\infty = 25$ minutes. *Upper panel:* Mass accretion rate relative to the Hoyle-Lyttleton analytic prediction (eq. [1]). *Middle panel:* Average specific angular momentum of the accreted material in units $R_a c_\infty$. Keplerian rotation at the surface of the object produces $j = 0.56$. *Lower panel:* Total time-integrated angular momentum accreted. In this low-resolution calculation, the instability does not reach full amplitude until $t \approx 22$. When a disk is present, j is near an extremum, switching sign when a disk forms rotating in the opposite direction. Large spikes in $\dot{M}$ occur when accretion disks dump their matter rapidly onto the accreting object.

Figure 2. Grayscale representation of the density in the flow at time t=10.14, from the calculation with uniform flow at moderate resolution. The accreting object lies at (0,0).

Figure 3. Accretion rates in the calculation with uniform flow and moderate resolution. Time units are $R_a/c_\infty = 25$min. *Upper panel:* Mass accretion rate relative to the Hoyle-Lyttleton analytic prediction (eq. [1]). *Middle panel:* Average specific angular momentum of the accreted material in units $R_a c_\infty$. Keplerian rotation at the surface of the object produces $j = 0.56$. *Lower panel:* Total time-integrated angular momentum accreted.

Figure 4. (a) Grayscale representation of the density in the flow near the accreting object at time t=8.09. (b) Velocity field at the same time. The longest arrow corresponds to the maximum velocity of $v = 18.6c_\infty$. (c) Grayscale representation of the density at time t=8.07. (d) Grayscale representation of the density at time t=8.02.

Figure 5. Accretion rates in the calculation with uniform flow and high resolution. Time units are $R_a/c_\infty = 25$min. *Upper panel:* Mass accretion rate relative to the Hoyle-Lyttleton analytic prediction (eq. [1]). *Middle panel:* Average specific angular momentum of the accreted material in units $R_a c_\infty$. Keplerian rotation at the surface of the object produces $j = 0.56$. *Lower panel:* Total time-integrated angular momentum accreted.

Figure 6. Accretion rates in the calculation with a small velocity gradient and moderate resolution. Time units are $R_a/c_\infty = 25$min. *Upper panel:* Mass accretion rate relative to the Hoyle-Lyttleton analytic prediction (eq. [1]). *Middle panel:* Average specific angular momentum of the accreted material in units $R_a c_\infty$. Keplerian rotation at the surface of the object produces $j = 0.56$. *Lower panel:* Total time-integrated angular momentum accreted.

Figure 7. Accretion rates in the calculation with a small density gradient and moderate resolution. Time units are $R_a/c_\infty = 25$min. *Upper panel:* Mass accretion rate relative to the Hoyle-Lyttleton analytic prediction (eq. [1]). *Middle panel:* Average specific angular momentum of the accreted material in units $R_a c_\infty$. Keplerian rotation at the surface of the object produces $j = 0.56$. *Lower panel:* Total time-integrated angular momentum accreted.

Figure 8. Accretion rates in the calculation with a moderate density gradient and moderate resolution. Time units are $R_a/c_\infty = 25$min. *Upper panel:* Mass accretion rate relative to the Hoyle-Lyttleton analytic prediction (eq. [1]). *Middle panel:* Average specific angular momentum of the accreted material in units $R_a c_\infty$. Keplerian rotation at the surface of the object produces $j = 0.56$. *Lower panel:* Total time-integrated angular momentum accreted.

Figure 9. Power spectral density (PSD) plots of the specific angular momentum accretion rates from two different calculations. Units of frequency are $c_\infty/R_a \approx 0.67\text{mHz}$. *Upper panel:* PSD of the specific angular momentum accretion rate from the calculation presented in §3.3.1 (small density gradient, moderate resolution grid). *Lower panel:* PSD of the specific angular momentum accretion rate from the calculation presented in §3.1.1 (uniform flow, low resolution grid). The peak at low frequency is caused by the flip-flop behavior, although it is not strong enough to be considered quasiperiodic. The reduction in power at high frequencies in the lower panel is due to the lower spatial resolution in the calculation.

Figure 10. Specific angular momentum of the accreted material plotted against the mass accretion rate for four different calculations in this paper. Each dot represents average quantities over 20 timesteps. *Upper left:* From the calculation in §3.2 with a velocity gradient. *Upper right:* From the calculation in §3.3.1 with a small density gradient. *Lower left:* From the calculation in §3.3.1 with a moderate density gradient. *Lower right:* From the calculation in §3.1.3 with uniform flow at high resolution.

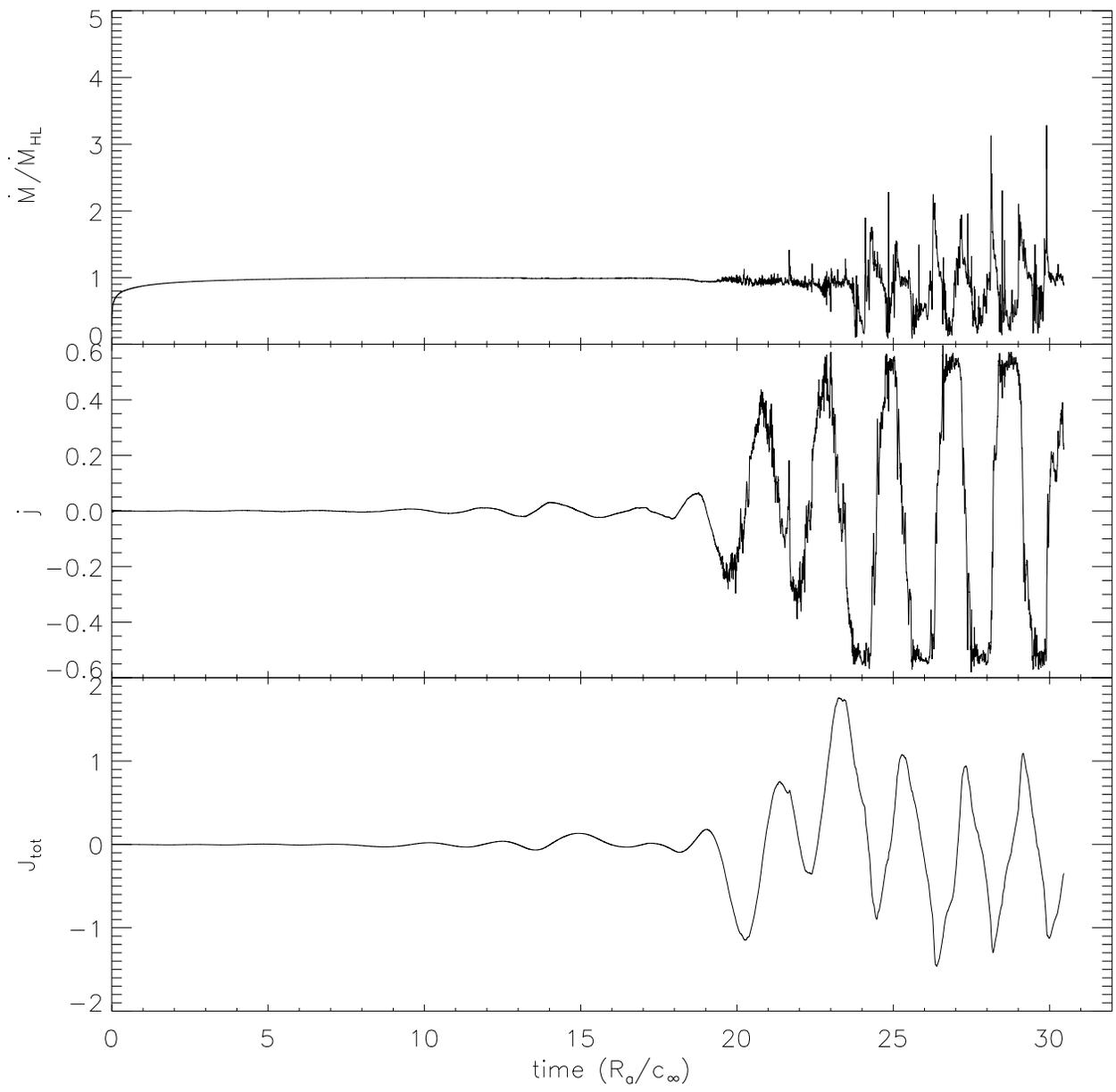

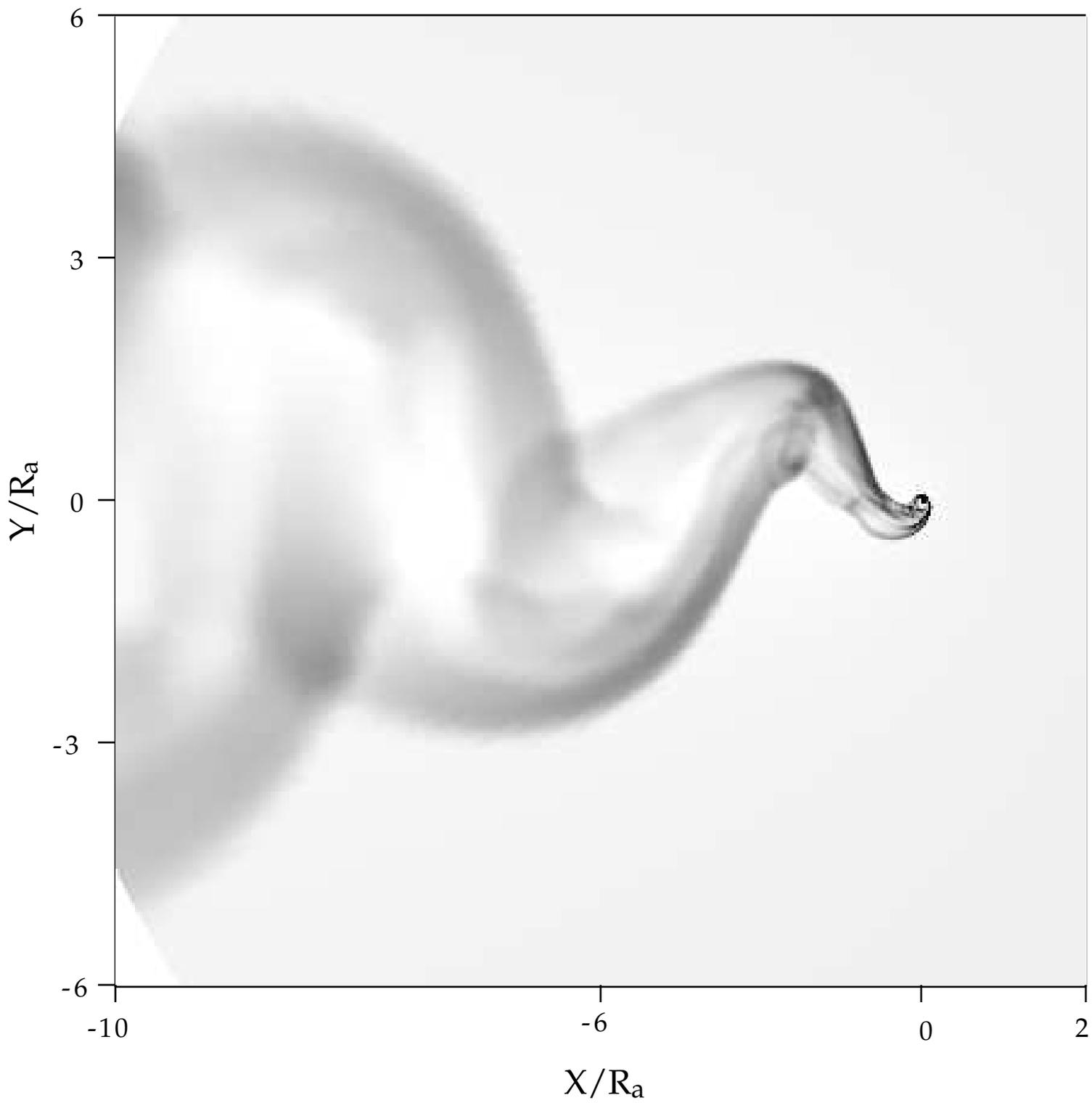

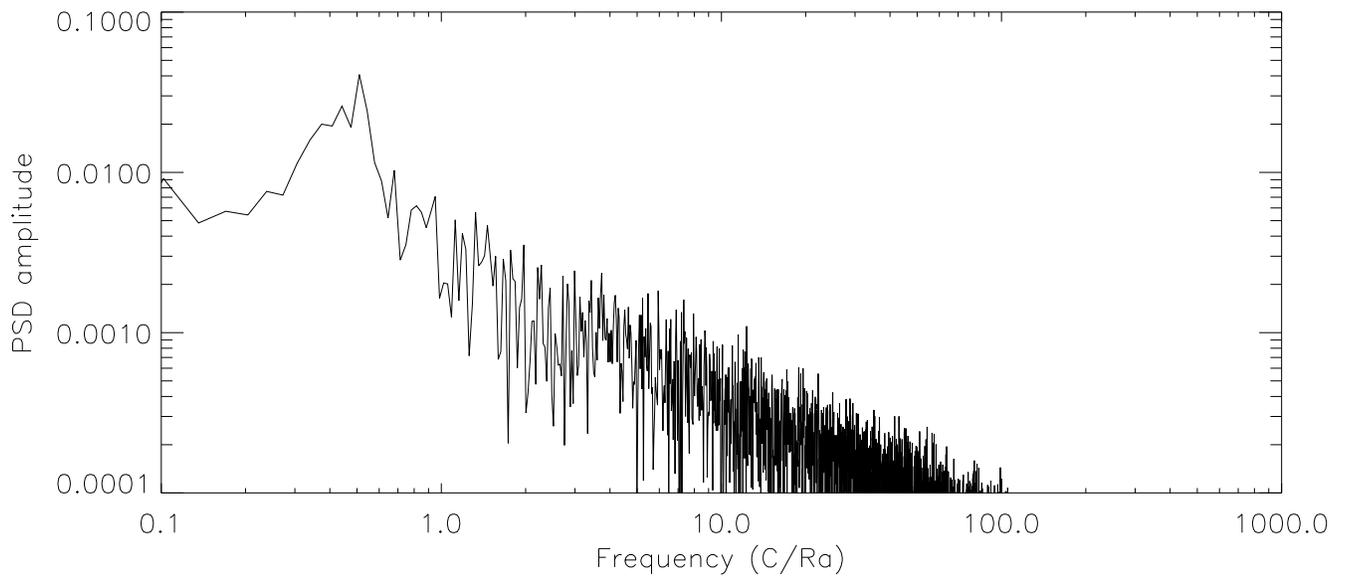
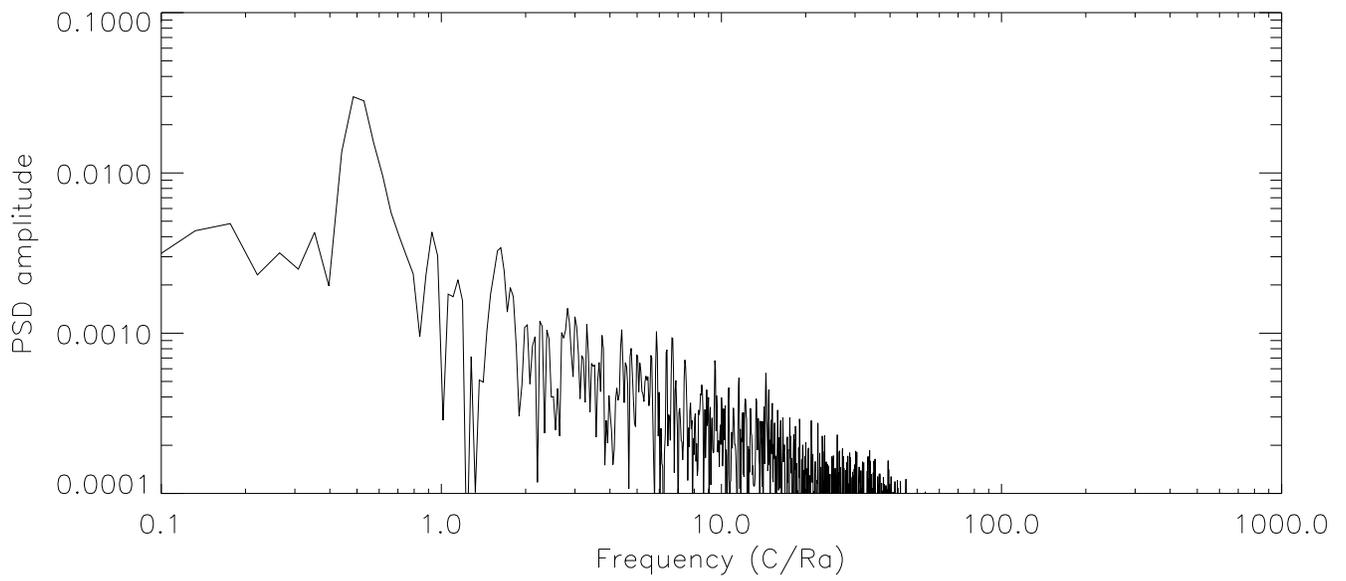

| System | $P_{orb}$ (days) | $v_w$ (km s$^{-1}$) | $\epsilon_\rho$ | $\epsilon_v$ |
|---|---|---|---|---|
| *O/B-type Systems* | | | | |
| Cen X-3 | 2.09 | 1,017 | $2.3 \times 10^{-2}$ | $4.5 \times 10^{-3}$ |
| OAO 1657-415 | 10.4 | 917 | $6.5 \times 10^{-3}$ | $5.5 \times 10^{-4}$ |
| Vela X-1 | 8.96 | 904 | $7.8 \times 10^{-3}$ | $7.5 \times 10^{-4}$ |
| 4U 1538-52 | 3.73 | 917 | $1.7 \times 10^{-2}$ | $2.9 \times 10^{-3}$ |
| GX 301-2 | 41.5 | 883 | $1.9 \times 10^{-3}$ | $6.8 \times 10^{-5}$ |
| *Be-type Systems* | | | | |
| 4U 0115+63 | 24.31 | 906 | $1.5 \times 10^{-3}$ | $7.4 \times 10^{-5}$ |
| EXO 2030+375 | 46.03 | 906 | $1.6 \times 10^{-3}$ | $5.1 \times 10^{-5}$ |
| 0535+26 | 110.58 | 926 | $6.1 \times 10^{-4}$ | $1.0 \times 10^{-5}$ |

Table 1: Expected values of density and velocity gradients in some typical high-mass X-ray binary systems, using data compiled by Chakrabarty (1996). The stellar wind velocity $v_w$ is computed as the escape speed at the surface of the star given its spectral type. Gradients are computed using equations (6), (7), (8), and (9) in the text.

| Mach | $\epsilon_\rho$ | $\epsilon_v$ | Resolution | $(N_r, N_\phi)$ | $t_{\text{final}}$ | $<\dot{M}>$ | $<P_{osc}>$ |
|---|---|---|---|---|---|---|---|
| 4 | 0.0 | 0.0 | Low | (200,100) | 30.5 | 0.94 | 1.89 |
| 4 | 0.0 | 0.0 | Moderate | (224,224) | 12.4 | 0.87 | 1.98 |
| 4 | 0.0 | 0.0 | High | (272,320) | 8.7 | 0.85 | 2.28 |
| 4 | 0.0 | 0.005 | Moderate | (224,224) | 31.1 | 0.83 | 1.96 |
| 4 | 0.005 | 0.0 | Moderate | (224,224) | 16.8 | 0.87 | 1.96 |
| 4 | 0.025 | 0.0 | Moderate | (224,224) | 18.6 | 0.86 | 1.96 |

Table 2: Calculations presented in this paper, their parameter values, and some results. Given for each calculation are: Mach number of the flow at infinity (the sound speed at infinity is always 250 km s$^{-1}$); size of the transverse density gradient $\epsilon_\rho$ (see eq. [8]); size of the transverse velocity gradient $\epsilon_v$ (see eq. [9]); a relative description of the spatial resolution employed; the grid size (number of radial zones $N_r$, number of azimuthal zones $N_\phi$); the final time to which the calculation is integrated $t_{\text{final}}$ (in units $R_a/c_\infty$); the average mass accretion rate found $<\dot{M}>$, relative to the Hoyle-Lyttelton prediction (eq. [1]); the average period of the flip-flop oscillations found $<P_{osc}>$ in units $R_a/c_\infty$.

1